\documentstyle[12pt,epsfig]{article} 

\begin{document}  
\vspace*{-2cm}  
\renewcommand{\thefootnote}{\fnsymbol{footnote}}  
\begin{flushright}  
hep-ph/0108221\\
PSI-PR-01-11\\  
Aug. 2001\\  
\end{flushright}  
\vskip 45pt  
\begin{center}  
{\Large \bf Resummation of angular dependent corrections in spontaneously
broken gauge theories} \\
\vspace{1.2cm} 
{\bf  
Michael Melles\footnote{Michael.Melles@psi.ch}   
}\\  

\begin{center}
Paul Scherrer Institute (PSI), CH-5232 Villigen, Switzerland. 
\end{center}

\vspace{20pt}  
\begin{abstract}
Recent investigations of electroweak radiative corrections have revealed the
importance of higher order contributions in high energy processes,
where the size of typical corrections can exceed those associated with QCD considerably. 
Beyond one loop,
only universal (angular independent) corrections are known to all orders except 
for massless $e^+ e^- \longrightarrow f {\overline f}$ processes where also angular dependent
corrections exist in the literature.
In this paper we present general arguments for the consistent
resummation of angular dependent subleading (SL) logarithmic corrections to all orders in 
the regime where all invariants are still large compared to the gauge boson masses.
We discuss soft isospin correlations, fermion mass and gauge boson
mass gap effects, the longitudinal 
and Higgs boson sector as well as mixing contributions including CKM effects 
for massive quarks. Two loop arguments
are interpreted in the context of the 
effective high energy effective theory based on the Standard Model Lagrangian
in the symmetric basis with the appropriate matching conditions to include
the soft QED regime. The result is expressed in exponentiated operator
form in a CKM-extended isospin space in the symmetric basis.
Thus, a full electroweak
SL treatment based on the infrared evolution equation method is
formulated for arbitrary high energy processes at future colliders. 
Comparisons with known results are presented.
\end{abstract}
\end{center}  
\vskip12pt

\setcounter{footnote}{0}  
\renewcommand{\thefootnote}{\arabic{footnote}}  
  
\vfill  
\clearpage  
\setcounter{page}{1}  
\pagestyle{plain} 
 
\section{Introduction} 

Future colliders in the TeV energy regime will attempt to clarify the physics responsible for
electroweak symmetry breaking. In this context it is important to understand the radiative corrections
from QCD as well as from the electroweak Standard Model (SM) sufficiently in order to disentangle
new physics effects. For precision measurements in the percentile regime, higher order 
{\it electroweak} radiative
corrections are crucial for this purpose \cite{habil}.

The exponentiation of electroweak double logarithms (DL) \cite{flmm} has now been established by
independent two loop calculations in the fermionic sector \cite{m2,bw,hkk} and therefore, using group
theoretical arguments and the equivalence theorem, also in the longitudinal sector as
was first derived in Ref. \cite{m1} via the infrared evolution equation method.

For universal, i.e. process independent subleading (SL) logarithms, the identity of the splitting function
approach and the physical fields calculation was shown in Refs. \cite{m1,m3} at
the one loop level. In Ref. \cite{dp} general formulae were presented
for arbitrary DL and SL corrections at one loop including process dependent angular terms.
The importance of these latter corrections was also discussed at one loop
in Ref. \cite{brv2}.
At higher orders, these corrections are important
and were first given for massless four fermion processes in Ref. \cite{kps}. For this particular group of
processes even sub-subleading corrections have been calculated by employing QCD results and
subtracting the QED corrections from the SM \cite{kmps}. In both cases the effects of 
higher order subleading terms are large and
need to be included in a consistent treatment.

In this paper, we are interested in calculating the angular dependent corrections to SL accuracy
to all orders for arbitrary electroweak processes. These include fermion mass terms and the associated
CKM mixing effects (which are absent for massless quarks since they are automatically mass eigenstates),
other mixing and mass gap effects of the electroweak gauge bosons and the longitudinal sector.
We assume that all invariants $2p_ip_j \gg M^2$, where $M$ denotes the gauge boson mass
with $M_Z \approx M_W \equiv M$.

Although we can not directly use the framework of QCD for the SM, at high energies we can
use a description based on the symmetric basis in which all terms with a mass dimension are
neglected \cite{habil}. This formulation implies for instance in the neutral scalar sector that
the relevant fields are $\phi_o$ and ${\phi_o}^*$. While strictly speaking we are calculating
corrections to amplitudes with these fields, the translation to the mass eigenstates is for
the most part straightforward. In our example we have the relations
\begin{equation}
H(x)=\frac{1}{\sqrt{2}} \left( {\phi_o}^*(x)+\phi_o(x) \right) \;\;,\; 
\chi(x)=\frac{i}{\sqrt{2}} \left( {\phi_o}^*(x)-\phi_o(x) \right)  
\end{equation}
Thus, the corrections factorize analogously for the mass eigenstates. Only in the neutral
transverse sector we can not directly assign a well defined isospin. In this case, 
and for corrections involving gauge bosons only, we
have to consider only the amplitude involving the non-Abelian field $W^3$ since the mixing
contribution with the $B$ field does not possess self interactions.

The contributions from the regime below the scale $M$ are due only to QED and 
can be incorporated via the appropriate matching conditions in the framework of the
infrared evolution equation method \cite{kl}.

In section \ref{sec:ga} we give general two loop arguments leading up to a factorization in
operator form for the angular dependent corrections in the physical basis.
In section \ref{sec:adc} the results of
section \ref{sec:ga} are generalized in the framework of the infrared evolution equation method
based on the Lagrangian in the symmetric basis and matching at the weak scale.
In section \ref{sec:comp}, the results in the effective description are compared at one loop 
with results in the physical basis in order to clarify the method, and to the massless
limit for four fermion processes also at higher orders.
We summarize our results in section \ref{sec:con}.

\section{General two loop arguments} \label{sec:ga}

Our general strategy in the following is to investigate if spontaneous symmetry breaking
effects can lead to novel angular dependent SL-corrections in comparison to the known
contributions in massless unbroken gauge theories at two loop order. As discussed below,
the results from the latter are used as input in our factorization for the electroweak
angular dependent corrections.

In the SM, the novel effects arising due to the electroweak symmetry breaking mechanism
are the mixing of the mass eigenstates, the existence of a scalar sector (which is not
mass suppressed at high energies) and the mass gap of the electroweak gauge bosons.
In addition to a massless non-Abelian theory
there are massive particles and in practice, these mass terms regularize
collinear divergences and are important for phenomenological predictions for future
colliders.

The effect of mixing is important for photon and Z-boson final states, since an external
$Z$-boson at one loop mixes with the photon field for instance.
However, these mixing terms
are not related to angular dependent corrections and are discussed in more
detail in Refs. \cite{dp,m1}.
A new complication is in principle given by the CKM-matrix elements, which enter in the
couplings of massive quarks to the charged
gauge bosons\footnote{CKM matrix elements also
occur in the scalar-quark-antiquark coupling. These diagrams, however, do not give rise
to SL angular terms.}. For universal corrections, using the conservation of the non-Abelian group charge,
these terms are of the form
\begin{equation}
\sum_{k=1}^3 V^*_{ik} V_{kj} f(m_i,m_k,m_j)
\end{equation}
where $f(0,0,0)=1$. For the exchange of the heavy gauge bosons, the mass terms
only lead to additional mass suppressed contributions if $m_l \leq M$.
The unitarity of the CKM matrix 
implies that the mass independent term in $f$ leads to the unit matrix.
For the general case with angular dependent corrections fermion mass terms remain, including 
in principle independent
CKM matrix elements, and thus, we have to convince ourselves
that at the $n$-loop level to SL accuracy no new 
$\log^{2n-1} \frac{s}{m^2_j} \log \frac{s}{t}$ arise from the CKM terms,
where $t$ denotes an invariant depending on the angle between the incoming and the outgoing
particle for instance.
These terms are discussed below but it should be noted that a massless calculation would not serve
as a check of these corrections since in this case they are automatically mass eigenstates.
For other final states, the mixing effects above the scale $M$
correspond to a change of basis, namely the symmetric basis, and therefore do not give
rise to new effects for physical cross sections.

\begin{figure}[t]
\centering
\epsfig{file=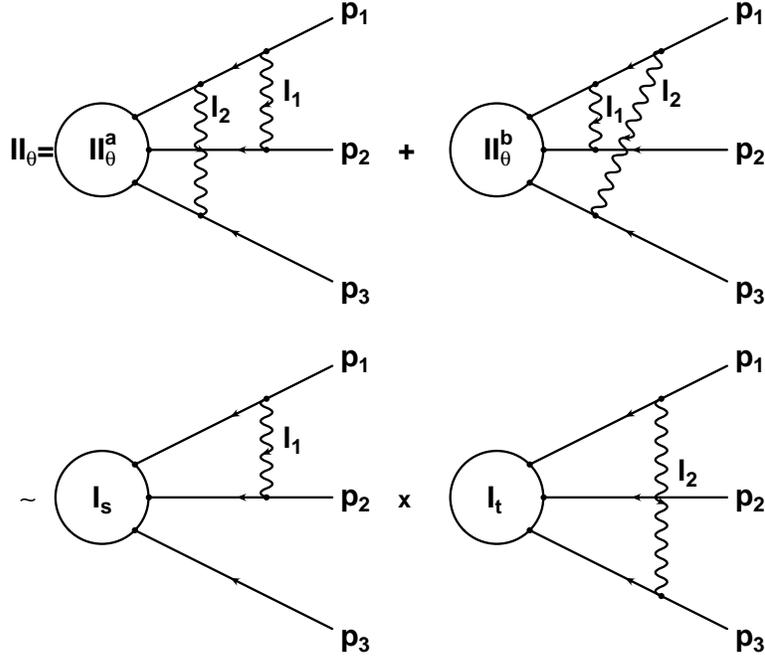,width=10cm}
\caption{Angular dependent two loop on-shell QED diagrams. The sum of II$^a_\theta$ and
II$^b_\theta$ factorizes in massive QED into the product of the two one loop
corrections (each with a different invariant and mass terms) in leading order.}
\label{fig:qedang}
\end{figure}

The new ingredient in spontaneously broken gauge theories are the mass terms. In the case
of the SM, the corrections from below the weak scale are only due to QED and the
angular SL terms in particular are only due to photon exchange. Thus, at the
two loop level, we have to consider corrections of the type depicted in Fig. \ref{fig:qedang}
and show that the mass terms do not lead to new effects compared to those in the
massless case.

The two scalar integrals of
Fig. \ref{fig:qedang}, regularized with a fictitious photon mass $\lambda$,
are given in massive QED by
\begin{eqnarray}
{\mbox{II}}^a_{\theta} &=& 4 s t \int \frac{d^4 l_1}{(2 \pi)^4} \int \frac{d^4 l_2}{(2 \pi)^4}
\left\{ \frac{1}{(l_1^2-\lambda^2)((p_1-l_1)^2-m_1^2)((p_1-l_1-l_2)^2-m_1^2)} \times 
\right. \nonumber \\
&& \left. \,\;\;\;\;\;\;\;\;\;\;\;\;\;\;\;\;\;\;\;\;\;\;\;\;\;\;\;\;\;\;\;\;\;
\frac{1}{((p_2+l_1)^2-m_2^2)
(l_2^2-\lambda^2)((p_3+l_2)^2-m_3^2)} \right\} \\ 
{\mbox{II}}^b_{\theta} &=& 4 s t \int \frac{d^4 l_1}{(2 \pi)^4} \int \frac{d^4 l_2}{(2 \pi)^4}
\left\{ \frac{1}{(l_1^2-\lambda^2)((p_1-l_2)^2-m_1^2)((p_1-l_1-l_2)^2-m_1^2)} \times 
\right. \nonumber \\
&& \left. \,\;\;\;\;\;\;\;\;\;\;\;\;\;\;\;\;\;\;\;\;\;\;\;\;\;\;\;\;\;\;\;\;\;
\frac{1}{((p_2+l_1)^2-m_2^2)
(l_2^2-\lambda^2)((p_3+l_2)^2-m_3^2)} \right\} 
\end{eqnarray}
denoting $s=2p_1p_2$, $t=2p_1p_3$ and 
where the $m_i$ are the masses of the external charged particles on their mass shell.
Thus, it is straightforward to see that the sum of the two diagrams factorizes to leading
order:
\begin{eqnarray}
{\mbox{II}}^a_{\theta}+{\mbox{II}}^b_{\theta} &=& 4 s t \int \frac{d^4 l_1}{(2 \pi)^4} \int 
\frac{d^4 l_2}{(2 \pi)^4}
\left\{ \frac{l_1^2+l_2^2-2p_1(l_1+l_2)}{(l_1^2-\lambda^2)((p_1-l_1)^2-m_1^2)((p_1-l_1-l_2)^2-m_1^2)} \times 
\right. \nonumber \\
&& \left. \,\;\;\;\;\;\;\;\;\;\;
\frac{1}{((p_2+l_1)^2-m_2^2)
(l_2^2-\lambda^2)((p_1-l_2)^2-m_1^2)((p_3+l_2)^2-m_3^2)} \right\} \nonumber \\ 
&\approx&  \int \frac{d^4 l_1}{(2 \pi)^4} \frac{2s}{(l_1^2-\lambda^2)
((p_1-l_1)^2-m_1^2)
((p_2+l_1)^2-m_2^2)} \times \nonumber \\
&& \int \frac{d^4 l_2}{(2 \pi)^4} \frac{2t}{(l_2^2-\lambda^2)((p_1-l_2)^2-m_1^2)
((p_3+l_2)^2-m_3^2)} 
\end{eqnarray}
The omitted cross term $2l_1l_2$ leads only to corrections containing three logarithms
at the two loop level. It is thus on the same level as the approximation in the beginning
of our discussion which only considers scalar integrals and can therefore be neglected.
To DL accuracy we can employ the Sudakov technique, parametrizing the loop momenta
along the external four momenta as
\begin{eqnarray}
l_1 &\equiv& v_1 \left( p_1 - \frac{m^2_1}{s} p_2 \right) + u_1 \left( p_2 - \frac{m^2_2}{s} p_1 \right)
+ {l_1}_\perp \\
l_2 &\equiv& v_2 \left( p_1 - \frac{m^2_1}{t} p_3 \right) + u_2 \left( p_3 - \frac{m^2_3}{t} p_1 \right)
+ {l_2}_\perp 
\end{eqnarray}
Thus, after rewriting the measure and integrating over the perpendicular components we find (omitting
the principle value parts):
\begin{eqnarray}
{\mbox{II}}^a_{\theta}+{\mbox{II}}^b_{\theta} &\sim& \frac{1}{8 \pi^2} \left[ \int^1_0 \frac{dv_1}{v_1}
\int^1_0 \frac{du_1}{u_1} \theta \left(su_1v_1-\lambda^2 \right) \theta \left(u_1-\frac{m_1^2}{s}
v_1 \right)  \theta \left(v_1-\frac{m_2^2}{s} u_1 \right) \right] \times \nonumber \\
&& \frac{1}{8 \pi^2} \left[ \int^1_0 \frac{dv_2}{v_2}
\int^1_0 \frac{du_2}{u_2} \theta \left(tu_2v_2-\lambda^2 \right) \theta \left(u_2-\frac{m_1^2}{t}
v_2 \right)  \theta \left(v_2-\frac{m_3^2}{t} u_2 \right) \right] \nonumber \\
&=& \frac{1}{8 \pi^2} \left[ \int^1_\frac{\lambda^2}{s} \frac{dv_1}{v_1} \int^1_\frac{\lambda^2}{sv_1} 
\frac{du_1}{u_1} - \int^\frac{\lambda m_2}{s}_\frac{\lambda^2}{s} \frac{dv_1}{v_1} \int^1_\frac{\lambda^2}{
s v_1} \frac{du_1}{u_1} - \int^\frac{m_2^2}{s}_\frac{\lambda m_2}{s} \frac{dv_1}{v_1} 
\int^1_\frac{s v_1}{m_2^2}
\frac{du_1}{u_1} \right. \nonumber \\
&& \;\;\;\;\;\;\;\;\; \left. - \int^\frac{\lambda m_1}{s}_\frac{\lambda^2}{s} \frac{du_1}{u_1} 
\int^1_\frac{\lambda^2}{
s u_1} \frac{dv_1}{v_1} - \int^\frac{m_1^2}{s}_\frac{\lambda m_1}{s} \frac{du_1}{u_1} 
\int^1_\frac{s u_1}{m_1^2}
\frac{dv_1}{v_1} \right] \times \nonumber \\
&& \frac{1}{8 \pi^2} \left[ \int^1_\frac{\lambda^2}{t} \frac{dv_2}{v_2} \int^1_\frac{\lambda^2}{tv_2} 
\frac{du_2}{u_2} - \int^\frac{\lambda m_3}{t}_\frac{\lambda^2}{t} \frac{dv_2}{v_2} \int^1_\frac{\lambda^2}{
t v_2} \frac{du_2}{u_2} - \int^\frac{m_3^2}{t}_\frac{\lambda m_3}{t} \frac{dv_2}{v_2} 
\int^1_\frac{t v_2}{m_3^2}
\frac{du_2}{u_2} \right. \nonumber \\
&& \;\;\;\;\;\;\;\;\; \left. - \int^\frac{\lambda m_1}{t}_\frac{\lambda^2}{t} \frac{du_2}{u_2} 
\int^1_\frac{\lambda^2}{
t u_2} \frac{dv_2}{v_2} - \int^\frac{m_1^2}{t}_\frac{\lambda m_1}{t} \frac{du_2}{u_2} 
\int^1_\frac{t u_2}{m_1^2}
\frac{dv_2}{v_2} \right] \nonumber \\
&=& \frac{1}{8 \pi^2} \left[ -\frac{1}{4} \log^2 \frac{s}{m_1^2} -\frac{1}{4} \log^2 \frac{s}{m_2}
+ \frac{1}{2} \log \frac{s}{m_1^2} \log \frac{s}{\lambda^2} + \frac{1}{2} \log \frac{
s}{m_2^2} \log \frac{s}{\lambda^2} \right] \times \nonumber \\
&& \frac{1}{8 \pi^2} \left[ -\frac{1}{4} \log^2 \frac{t}{m_1^2} -\frac{1}{4} \log^2 \frac{t}{m_3}
+ \frac{1}{2} \log \frac{t}{m_1^2} \log \frac{t}{\lambda^2} + \frac{1}{2} \log \frac{
t}{m_3^2} \log \frac{t}{\lambda^2} \right] \label{eq:qedang} 
\end{eqnarray}
The important point about the result in Eq. (\ref{eq:qedang}) is not only the factorized form in terms of
the two massive one loop form factors but also the fact that the fermion mass terms correspond to
each external on shell line in the amplitude. Thus, by rewriting the term in the bracket of the last
line in Eq. (\ref{eq:qedang}) as
\begin{eqnarray}
&& -\frac{1}{4} \log^2 \frac{t}{m_1^2} -\frac{1}{4} \log^2 \frac{t}{m_3}
+ \frac{1}{2} \log \frac{t}{m_1^2} \log \frac{t}{\lambda^2} + \frac{1}{2} \log \frac{
t}{m_3^2} \log \frac{t}{\lambda^2}  \nonumber \\ &=& \frac{1}{2}
\log^2 \frac{t}{\lambda^2}-\frac{1}{4} \left( \log^2 \frac{m^2_1}{\lambda^2} + \log^2 \frac{m^2_3}{\lambda^2}
\right) \nonumber \\ &\approx& \frac{1}{2} \log^2 \frac{s}{\lambda^2} + \log \frac{s}{\lambda^2} 
\log \frac{t}{s} 
-\frac{1}{4} \left( \log^2 \frac{m^2_1}{\lambda^2} + \log^2 \frac{m^2_3}{\lambda^2} \right)
\end{eqnarray}
we see that the SL angular terms are indeed independent of the fermion mass terms.
For the corrections involving the invariant $u\equiv 2 p_2p_3$
the situation is analogous.

This type of factorization can be generalized on theoretical grounds to the situation in the
SM. It should be noted that all fermion mass singularities in the SM only arise through
photon radiation or coupling renormalization. The latter is not important in our discussion here and
is anyhow sub-subleading at higher orders. 
The exchange of the heavy gauge bosons does not lead to fermion mass singular terms assuming that
all $m_i \leq M$.

\begin{figure}
\centering
\epsfig{file=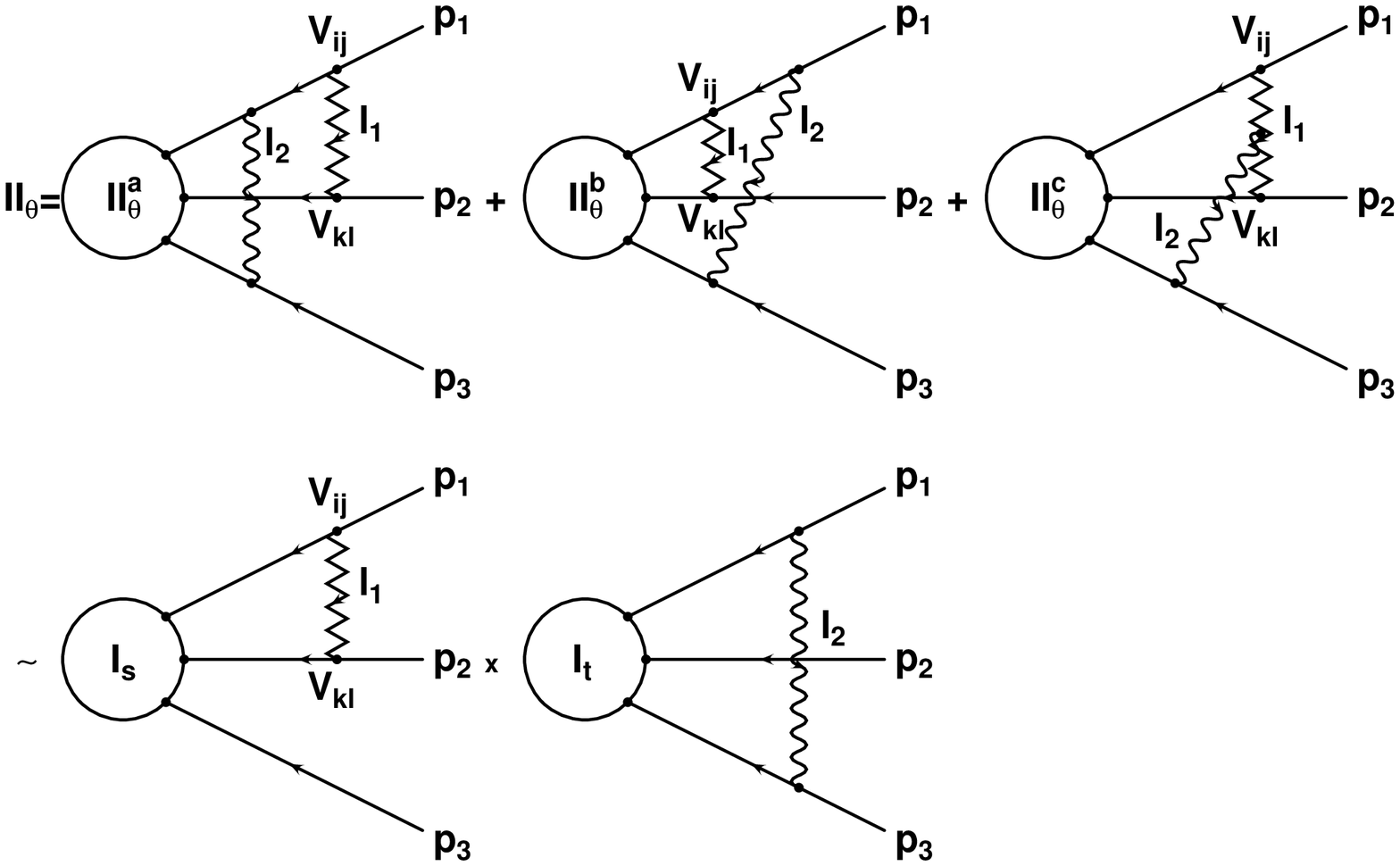,width=11cm} \\
\vspace{1cm}
\epsfig{file=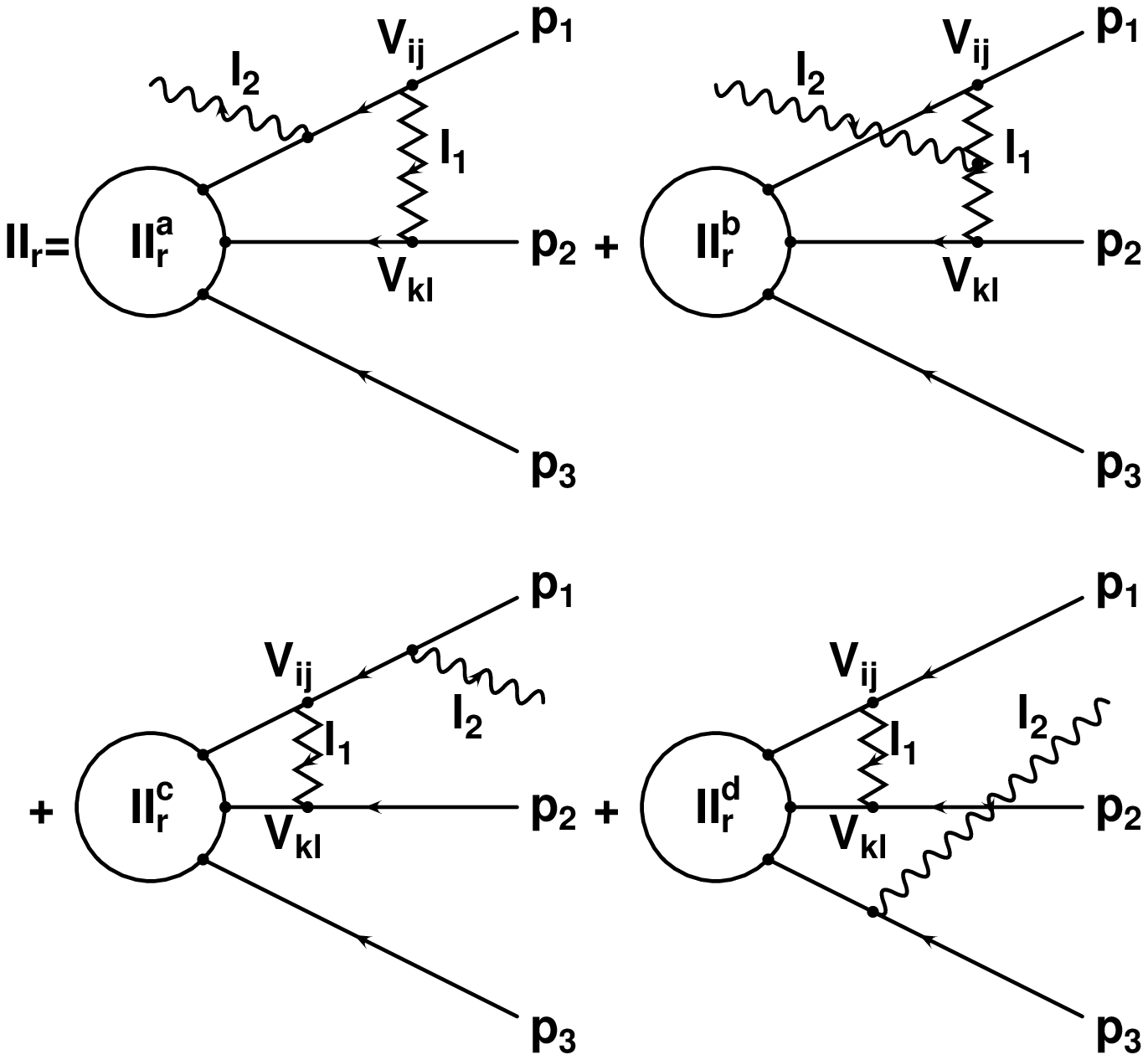,width=7cm}
\caption{Electroweak angular dependent two loop on-shell diagrams involving CKM matrix elements
(denoted by $V_{ij}$). 
The wavy line denotes a photon, the zigzag line a $W^\pm$.
The sum of the virtual diagrams II$^a_\theta$, II$^b_\theta$ and
II$^c_\theta$ must factorize for the leading mass singular terms into the product of the two one loop
corrections (each with a different invariant and mass terms). The formal reason for this factorization
is provided by the fact that the real emission diagrams II$^a_r$ and II$^b_r$ are free of fermion mass
singular terms due to the off-shellness generated by the $W^\pm$ exchange. Thus only the factorized 
real corrections of diagrams II$^c_r$ and II$^d_r$ contribute fermion mass singular terms and the
KLN theorem then yields the analogous factorization for the above virtual corrections.
For the full SM corrections, however, one must include all diagrams since the photon
cannot be separated from the other gauge bosons in a gauge invariant way.}
\label{fig:ckm}
\end{figure}

In order to generalize the QED type factorization derived in Eq. (\ref{eq:qedang})
at the two loop level for diagrams involving fermion mass singularities, 
one has to consider in particular 
corrections such as the ones depicted in Fig. \ref{fig:ckm}.
Compared to QCD, these are novel type diagrams involving mixing effects in the
fermionic sector through CKM matrix elements.
The required factorized form shown there for the virtual diagrams\footnote{Note that here 
we only
consider the fermion mass singular terms. For the full SM factorization of SL-logarithmic terms,
all diagrams need to be calculated since the photon cannot be separated from the other
gauge bosons in a gauge invariant way.}, is actually
a consequence of the type of fermion mass singular terms which occur in the associated real emission diagrams
shown in the lower half of Fig. \ref{fig:ckm}. The diagrams II$^a_r$ and II$^b_r$ do not produce fermion
mass singular terms due to the virtuality ($\geq M$) of the line emitting the photon (wavy line). Thus, only
the real emission diagrams in the last row remain which are of the required factorized form.

The connection between the real and virtual diagrams is now provided by the KLN theorem \cite{k,ln}.
It states that, as a consequence of unitarity, transition probabilities are free of mass singularities
when summed over all degenerate final states. This is true order by order in 
perturbation theory in renormalization schemes which do not introduce mass singular terms via the
renormalization constants. As mentioned above, we are not concerned about the last point here.

Thus, the only way to cancel the real emission fermion mass singularities in Fig. \ref{fig:ckm}
in fully inclusive cross sections is by the factorization of the associated virtual ones as depicted
in the figure. As mentioned above, the exchange of the heavy gauge bosons does not lead to fermion
mass singular terms. Taking into account all diagrams of the SM 
leads, however, to uncanceled $\log^{2n} \frac{s}{M^2}$ 
Bloch-Nordsieck violating terms in fully inclusive
cross sections since the initial state carries isospin \cite{ccc1}.

The scalar sector in itself, after the application
of the equivalence theorem, poses no new problem since the angular SL-terms originate from
soft corrections. In Refs. \cite{dp,m3} the universal nature of Yukawa enhanced SL corrections was
investigated and all higher order SL terms resummed in Ref. \cite{m3}. By direct inspection of the
full electroweak Feynman rules it can easily be seen that there are no angular dependent
SL corrections with Yukawa terms possible which are not mass suppressed.
Thus, whether we consider fundamental fermions or scalars charged under
the unbroken gauge group leads to
the same form of factorized corrections. 

A more serious problem, in principle, is given by the mass gap of the gauge bosons, since
there could be novel soft angular corrections compared to the equal mass case. This, however,
is not the case since the soft angular terms from virtual scattering amplitudes in the
full SM are fixed by the pure real QED corrections in the regime where the experimental
energy resolution $\Delta E$ is below the weak scale $M$. 
This energy cut provides a well defined observable in the SM without the inclusion of
real heavy gauge boson emission.
Since the real QED corrections are always
proportional to $\log \frac{2 \Delta E}{\lambda}$ at one loop, and
exponentiates at higher orders, the infrared finiteness of observable cross sections
fixes the type of soft virtual angular corrections to that of pure QED effects.

The general structure of pole terms for virtual two loop massless QCD scattering amplitudes
was presented in Ref. \cite{cat} and subsequently confirmed by explicit
two loop calculations in Refs. \cite{dix,glov} in two to two processes.
While the result presented in \cite{cat} has been formulated for $\overline{\mbox{MS}}$
renormalized on-shell amplitudes, the form of the factorization is regularization
scheme independent at least up to single pole terms. An application of the factorization
formula in the QED limit to elastic Bhabha scattering
was presented in Ref. \cite{gtb}. It uses mass regulators
for the soft and collinear divergences. To SL accuracy in QED, one loop results 
calculated in this mass regulator scheme determine
the corresponding two loop logarithmic corrections.

Also in the general case, a
main feature of the factorization formula at the two loop level is that - up to terms
corresponding to higher order running coupling terms - all leading and subleading
pole terms are determined by the one loop divergences in color space. The latter
was formulated in Ref. \cite{sey} and serves as a convenient way to incorporate
soft color correlations in QCD scattering amplitudes. These are novel effects in
non-Abelian gauge theories and need to be properly included.

In the electroweak theory, the corresponding correlations are due to
the exchange of the $W^\pm$ and--in the neutral sector--the $Z$ gauge
bosons only, since the photon 
does not change the flavor. Thus, it is legitimate for these terms to only
consider a theory above the scale M, i.e. an unbroken massless SU(2)$\times$U(1) theory
with all gauge bosons regularized by the same mass $M$.
This theory then has angular corrections in the same form as QCD, now however, formulated
in the $n$-particle space. 

From the above arguments it is now clear that to SL accuracy, neglecting RG-SL terms for now\footnote{
These terms are discussed in Ref. \cite{m4}, and are related to external lines only. Thus they do not depend
on angular terms and are therefore not important for the current discussion. The exponentiation of the
universal terms was discussed in Refs. \cite{flmm,m1,m3}.},
the virtual SM corrections in the on-shell scheme at the one and 
two loop level
for processes with n external lines
can be expressed as 
\begin{eqnarray}
{\cal M}^{i_1,...,i_n}_{(1)} (p_1,...,p_n;m_1,...,m_n;M,\lambda) \!\!\!&=&\!\!\! 
I^{(1)}_n \left( \{ p_k \}, \{m_l \},M,\lambda \right) \! \times \nonumber \\ && \!\!\!
{\cal M}^{i_1,...,i_k^\prime,...,i^\prime_l,...,i_n}_{\rm Born} (p_1,...,p_n;m_1,...,m_n) 
\nonumber \\
{\cal M}^{i_1,...,i_n}_{(2)} (p_1,...,p_n;m_1,...,m_n;M,\lambda) \!\!\!&=&\!\!\! \frac{1}{2} 
I^{(1)}_n \left( \{ p_k \}, \{m_l \},M,\lambda \right) I^{(1)}_n \left( \{ p_k \}, \{m_l \}
,M,\lambda \right) 
\! \times \nonumber \\ && \!\!\!
{\cal M}^{i_1,...,i_k^\prime,...,i^\prime_l,...,i_n}_{\rm Born} (p_1,...,p_n;m_1,...,m_n)
\label{eq:tlfac}
\end{eqnarray}
All finite terms at one loop and the corresponding terms at two loops below SL accuracy are omitted in
Eq. (\ref{eq:tlfac}).
The $n$-particle space operator $I^{(1)}_n \left( \{ p_k \}, \{m_l \},M,\lambda \right)$ 
is determined by the one loop
structure and its explicit form is given by:
\begin{eqnarray}
&& \!\!\!\!\!\!\!\!\!\! I^{(1)}_n \left( \{ p_k \}, \{m_l \} ,M,\lambda \right) = 
\frac{e^2}{16 \pi^2} \sum^n_{k=1} \left\{ - \frac{1}{2} \left[
C^{\rm ew}_{i^\prime_k,i_k} \log^2 \frac{s}{M^2} + \delta_{i^\prime_k,i_k} Q_k^2 \left( 2 \log \frac{s}{M^2}
\log \frac{M^2}{\lambda^2} + \right. \right. \right. \nonumber \\ && \!\!\!\!\!\!\!\!\!\!
\left. \left. \left. \log^2 \frac{M^2}{\lambda^2}-\log^2 \frac{m_k^2}{\lambda^2} \right) \right] + 
\delta_{i^\prime_k,i_k}^{ \rm SL} \log \frac{s}{M^2} + 
2 \sum^n_{l < k} \sum_{V_a=A,Z,W^\pm} \!\!\!\!
I^{V_a}_{i^\prime_k,i_k} I^{
{\overline V}_a}_{i^\prime_l,
i_l} \log \frac{s}{m^2_{V_a}} \log \frac{2 p_lp_k}{s} \right\} \label{eq:olo}
\end{eqnarray}
where all lines are assumed as incoming,
the symbols $I^{V_a}_{i^\prime_k,i_k}$ denote the couplings of the
${\overline \varphi}_{i_k} V_a \varphi_{i^\prime_k}$ vertices and $C^{\rm ew}_{i^\prime_k,i_k}$ the corresponding
electroweak eigenvalue of the Casimir operator. The DL corrections were first obtained in Ref.
\cite{flmm} and for longitudinal polarizations and Higgs final states in Ref. \cite{m1}, in both
cases actually to all orders.
The general form of the angular dependent corrections at one loop was first derived in Ref. \cite{dp}.
The subleading universal corrections $\delta_{i^\prime_k,i_k}^{ \rm SL}$ depend on the external 
line only and are
given in Refs. \cite{dp,m1,m3}.

In the non-diagonal part of Eq. (\ref{eq:olo}) it can be seen that in distinction to an unbroken
gauge theory, the angular terms include quark mixing effects
included in the couplings
$I^{V_a}_{i^\prime_k,i_k}$) and depend on different masses, i.e. the masses of the electroweak gauge
bosons. Due to our discussion above, however, fermion mass singularities are fixed by
the KLN theorem and the soft angular terms of QED origin by the
infrared finiteness of semi inclusive cross sections.

\section{Angular dependent corrections in the effective theory} \label{sec:adc} 

In this section we generalize the two loop results of the previous section in Eq. (\ref{eq:tlfac})
to all orders by reinterpreting the angular terms in the language of the high energy effective theory 
based on
the Lagrangian in the symmetric basis. The physical picture is that of the approximately restored
SU(2)$\times$U(1) gauge symmetry at high energies where terms with a mass dimension can be neglected.
The soft QED effects below the scale $M$, set by the electroweak gauge bosons, can be included via
matching at that scale. For a thorough discussion see Ref. \cite{habil}.

Amplitudes in terms of the physical fields ($f$) are given in terms of
superpositions of the fields in the unbroken phase ($u$) as follows \cite{habil}:
\begin{equation}
{\cal M}^{f_1,...,f_n}(\{p_{k}\};\{m_{l}\};M,\lambda) =\sum_{u_1,...,u_n }
\prod_{j=1}^n C^{f_ju_j}
{\cal M}^{u_1,...,u_n}(\{p_{k}\};\{m_l\};M,\lambda) \label{eq:lc}
\end{equation}
where the $C^{f_ju_j}$ denote the corresponding mixing coefficients.
We note that at one loop, also the renormalization conditions involving the mixing
coefficients need to be included properly as discussed in Refs. \cite{m1,dp}.
The fields $u$ have a well defined isospin, but for angular dependent terms involving
CKM mixing effects, one has to include the extended isospin mixing appropriately in 
the corresponding couplings ${\tilde I}^{V_a}_{i^\prime_k,i_k}$. 
In the following we give the corrections only for
the amplitudes ${\cal M}^{u_1,...,u_n}(\{p_{k}\};\{m_l\};M,\lambda)$.

As far as the angular dependent terms are concerned, it is clear that above the scale $M$, i.e.
for a photon with mass $M$ rather than $\lambda$, the difference to the corresponding result in
Eq. (\ref{eq:tlfac}) is given by a change of basis, i.e.
\begin{eqnarray}
&& \sum_{k=1}^n \sum^n_{l < k} \sum_{V_a=A,Z,W^\pm}\!\!\!\!\!\! e^2I^{V_a}_{i^\prime_k,i_k} I^{
{\overline V}_a}_{i^\prime_l,
i_l} \log \frac{s}{M^2} \log \frac{2 p_lp_k}{s}  
{\cal M}^{f_{i_1},...,f_{i^\prime_k},...,f_{i^\prime_l},...,f_{i_n}}_{\rm Born}(\{p_{k}\};\{m_{l}\}) =
\nonumber \\ && \sum_{u_1,...u_n } \prod_{j=1}^n C^{f_ju_j} 
\sum_{k=1}^n \sum^n_{l < k} \sum_{V_a=B,W^a} \!\!\! {\tilde I}^{V_a}_{i^\prime_k,i_k} {\tilde I}^{
{\overline V}_a}_{i^\prime_l,
i_l} \log \frac{s}{M^2} \log \frac{2 p_lp_k}{s} \times \nonumber \\ && 
{\cal M}^{u_{i_1},...,u_{i^\prime_k},...,u_{i^\prime_l},...,u_{i_n}}_{\rm Born}(\{p_{k}\};\{m_{l}\})
\label{eq:angM}
\end{eqnarray}
The remaining terms are then just given by QED corrections of the type
\begin{equation}
\sum_{k=1}^n \sum^n_{l < k} Q_k Q_l
\log \frac{M^2}{\lambda^2} \log \frac{2 p_lp_k}{s} \label{eq:angl} 
\end{equation}
The terms in Eq. (\ref{eq:angl}) therefore correspond precisely to matching terms in analogy
to the situation for the universal SL logarithms \cite{m1,m3}. Thus, if ${\cal L}_{symm}$ is a valid
effective theory at high energies in the sense that it contains all relevant physical degrees
of freedom\footnote{For the angular dependent corrections in the massive quark sector
we must in principle 
include the CKM terms in the corresponding couplings as mentioned above.}, 
then matching at the scale $M$ must yield the corresponding
soft QED terms since these are fixed by the real QED corrections and their higher order behavior.

The arguments given in section \ref{sec:ga} with respect to the (non-)influence of spontaneous
symmetry breaking on the form of the
SL angular dependent corrections at the two loop level can easily be generalized
to arbitrary order in perturbation theory.
The associated exponentiation of the non-Abelian high energy regime is now in matrix form analogous
to QCD \cite{cat} and reads for the purely virtual corrections in the on-shell scheme
as follows\footnote{In this paper
we use a photon mass $\lambda$ regulator for the soft QED effects in order to simplify the
comparisons with calculations using the physical fields. This is identical to using a cutoff
in the exchanged perpendicular components of the emitted particles if the cutoff is much smaller
than all other parameters in the theory. If one wants to check the validity of the matching
conditions one has to use the general form given in Ref. \cite{habil}.}:
\begin{eqnarray}
&& \!\!\!\!\!\!\!\!\! {\cal M}_{\rm SL}^{u_{i_1},...,
u_{i_n}} \left( \{ p_k \}; \{m_l \}; M, \lambda \right) \rangle = 
\exp \left\{ - \frac{1}{2} \sum^{n_g}_{i=1} W^{\rm RG}_{g_i} (s,M^2)
 - \frac{1}{2} \sum^{n_f}_{i=1} W^{\rm RG}_{f_i} (s,M^2) \right. \nonumber \\ && \!\!\!\!\!\!\!\!\! \left.
 - \frac{1}{2} \sum^{n_\phi}_{i=1} W^{\rm RG}_{\phi_i} (s,M^2)+
 \frac{1}{8 \pi^2} \sum_{k=1}^n \sum^n_{l < k} \sum_{V_a=B,W^a} \!\!\! {\tilde I}^{V_a}_{i^\prime_k,i_k} 
 {\tilde I}^{
 {\overline V}_a}_{i^\prime_l,
 i_l} \log \frac{s}{M^2} \log \frac{2 p_lp_k}{s} 
 \right\} \nonumber \\
 && \!\!\!\!\!\!\!\!\! \times \exp \left[ - \frac{1}{2} \sum_{i=1}^{n_f} \left( w^{\rm RG}_{f_i}(s,\lambda^2)
 - w^{\rm RG}_{f_i}(s,M^2) \right)
 - \frac{1}{2} \sum_{i=1}^{n_{\rm w}} \left( w^{\rm RG}_{{\rm w}_i}(s,\lambda^2)
 - w^{\rm RG}_{{\rm w}_i}(s,M^2) \right) \right. \nonumber \\
 && \!\!\!\!\!\!\!\!\! \left. - \frac{1}{2} \sum_{i=1}^{n_\gamma} w_{\gamma_i}(M^2,m_j^2)
 + \sum_{k=1}^n \sum^n_{l < k} \left( w^\theta_{kl} \left(s,\lambda^2 \right) - 
w^\theta_{kl} \left(s,M^2 \right) \right)
 \right] \times \nonumber \\ &&
{\cal M}^{u_{i_1},...,u_{i^\prime_k},...,u_{i^\prime_l},...,u_{i_n}}_{\rm Born}
(\{p_{k}\};\{m_{l}\})
 \label{eq:angr}
\end{eqnarray}
where $n_g$ denotes the number of external gauge bosons (in the symmetric basis), $n_f$ the number
of external fermions and $n_\phi$ the number of external scalars (including Higgs particles).
The notation used in the regime below the scale $M$ is analogous.
The diagonal terms do not involve (CKM-extended) isospin rotated Born matrix elements. 
These occur only from the angular
terms above the scale $M$. 
At the two loop level, Eq. (\ref{eq:angr}) reproduces
Eq. (\ref{eq:tlfac}) after using Eq. (\ref{eq:lc}) 
up to RG-SL terms which we included here. The explicit expressions for the various
ingredients given in Eq. (\ref{eq:angr}) are given by
\begin{eqnarray}
 W^{\rm RG}_{\phi_i}(s,M^2)\!\!\!\!&=&\!\!\!\! \frac{\alpha(M^2)
  T_i(T_i+1)}{2 \pi } \left\{ \frac{1}{c} \log \frac{s}{M^2}
  \left( \log \frac{\alpha(M^2)}{\alpha
  (s)} - 1 \right) + \frac{1}{c^2}
  \log \frac{\alpha(M^2)}{\alpha(s)} \right\} \nonumber \\
  \!\!\!\!&&\!\!\!\!\!\!\!\!\!\!\!\!\!\!\!\!\!\!\!\! +\frac{\alpha^\prime(M^2) Y^2_i}{8 \pi } \left\{ \frac{1}{
  c^\prime} \log \frac{s}{M^2}
  \left( \log \frac{\alpha^\prime(M^2)}{\alpha^\prime
  (s)} - 1 \right) + \frac{1}{{c^\prime}^2}
  \log \frac{\alpha^\prime(M^2)}{\alpha^\prime(s)} \right\} \nonumber \\
  \!\!\!\!&&\!\!\!\!\!\!\!\!\!\!\!\!\!\!\!\!\!\!\!\! - \left[ \left( \frac{ \alpha(M^2)}{4 \pi}  T_i(T_i+1)+
  \frac{ \alpha^\prime(M^2)}{4 \pi} \frac{Y^2_i}{4} \right)  4 \log \frac{s}{M^2}
  - \frac{3}{2} \frac{ \alpha(M^2)}{4 \pi} \frac{m^2_t}{M^2} \log \frac{s}{m_t^2} \right] \label{eq:WpRG}
   \end{eqnarray}
   where $T_i$ denotes the total weak isospin of particle $i$ and $Y_i$ its hypercharge.
   Furthermore, we denote the $SU(2)$ coupling by $\alpha=\frac{g^2}{4 \pi}$ and that of $U(1)$ by
   $\alpha^\prime=\frac{{g^\prime}^2}{4 \pi}$.
   The two one loop contributions to the respective $\beta$-functions are given by
   $\beta_0=\frac{11}{12}C_A - \frac{1}{3}n_{gen}-\frac{1}{24}n_{h}$ and
   $\beta^\prime_0= - \frac{5}{9}n_{gen} -\frac{1}{24}n_{h}$, where $n_{gen}$ denotes the number
   of generations and $n_h$ the number of Higgs doublets.
   We can use $m_t$ in the logarithmic argument of the Yukawa enhanced correction \cite{m3}. In addition
   we denote
   $c=\frac{g^2}{4 \pi^2} \beta_0$ and $c^\prime=\frac{{g^\prime}^2}{4 \pi^2} \beta^\prime_0$.
   Analogously for fermions we have:
   \begin{eqnarray}
   W^{\rm RG}_{f_i}(s,M^2)&=& \frac{\alpha(M^2)
    T_i(T_i+1)}{2 \pi } \left\{ \frac{1}{c} \log \frac{s}{M^2}
    \left( \log \frac{\alpha(M^2)}{\alpha
    (s)} - 1 \right) + \frac{1}{c^2}
    \log \frac{\alpha(M^2)}{\alpha(s)} \right\} \nonumber \\
    && +\frac{\alpha^\prime(M^2) Y^2_i}{8 \pi } \left\{ \frac{1}{c^\prime} \log \frac{s}{M^2}
    \left( \log \frac{\alpha^\prime(M^2)}{\alpha^\prime
    (s)} - 1 \right) + \frac{1}{{c^\prime}^2}
    \log \frac{\alpha^\prime(M^2)}{\alpha^\prime(s)} \right\} \nonumber \\
    && - \left[ \left( \frac{ \alpha(M^2)}{4 \pi} T_i(T_i+1)+
    \frac{ \alpha^\prime(M^2)}{4 \pi}\frac{Y^2_i}{4} \right)  3 \log \frac{s}{M^2}
    \right. \nonumber \\ && \left.
    - \frac{ \alpha(M^2)}{4 \pi} \left( \frac{1+\delta_{i,{\rm R}}}{4} \frac{m^2_i}{M^2} + \delta_{i,{\rm L}}
    \frac{m^2_{i^\prime}}{4 M^2} \right)
    \log \frac{s}{m_t^2} \right] \label{eq:WfRG}
    \end{eqnarray}
    The last term contributes only for left handed bottom as well as for top quarks and
    $f^\prime$ denotes the corresponding isospin partner for left handed fermions.
    \begin{eqnarray}
    &&  W^{\rm RG}_{g_i}(s,M^2)= \frac{\alpha(M^2)
     T_i(T_i+1)}{2 \pi } \left\{ \frac{1}{c} \log \frac{s}{M^2}
     \left( \log \frac{\alpha(M^2)}{\alpha
     (s)} - 1 \right) + \frac{1}{c^2}
     \log \frac{\alpha(M^2)}{\alpha(s)} \right\} \nonumber \\
     && +\frac{\alpha^\prime(M^2) Y^2_i}{8 \pi } \left\{ \frac{1}{c^\prime} \log \frac{s}{M^2}
     \left( \log \frac{\alpha^\prime(M^2)}{\alpha^\prime
     (s)} - 1 \right) + \frac{1}{{c^\prime}^2}
     \log \frac{\alpha^\prime(M^2)}{\alpha^\prime(s)} \right\} \nonumber \\
     &&- \left( \delta_{i,{\rm W}} \frac{\alpha(M^2)}{\pi} \beta_0 + \delta_{i,{\rm B}}
     \frac{\alpha^\prime(M^2)}{\pi} \beta^\prime_0 \right) \log \frac{s}{M^2} \label{eq:WgRG}
     \end{eqnarray}
     We note that for external photon and Z-boson states we must include
     the mixing appropriately as discussed in Ref. \cite{m1}.
     For the terms entering from contributions below the weak scale we have for fermions:
     \begin{eqnarray}
     w^{\rm RG}_{f_i}(s,\lambda^2) & = &
     \frac{e_i^2}{8 \pi^2 } \left\{ \frac{1}{c} \log \frac{s}{m^2}
     \left( \log \frac{e^2(\lambda^2)}{e^2
     (s)} - 1 \right) -\frac{3}{2} \log \frac{s}{m^2} - \log \frac{ m^2}{\lambda^2} \right. \nonumber \\
     &&\left. + \frac{1}{c^2}
     \log \frac{e^2(m^2)}{e^2(s)} \left( 1- \frac{1}{3}\frac{e^2}{4 \pi^2}
     \sum_{j=1}^{n_f } Q_j^2 N^j_C \log \frac{m^2}{m_j^2}
     \right) \right\}
     \end{eqnarray}
     where $c=-\frac{1}{3}\frac{e^2}{4 \pi^2} \sum^{n_f}_{j=1}Q^2_j N_C^j$ and
     the running QED coupling
     given in Eq. (\ref{eq:aeff}) and $e_i\equiv e Q_i$.
     Analogously, for external W-bosons and photons we find:
     \begin{eqnarray}
     w^{\rm RG}_{{\rm w}_i}(s,\lambda^2) &=&
     \frac{e_i^2}{8 \pi^2 } \left\{ \frac{1}{c} \log \frac{s}{M^2}
     \left( \log \frac{e^2(\lambda^2)}{e^2
     (s)} - 1 \right) - \log \frac{ M^2}{\lambda^2} \right. \nonumber \\
     && \left. + \frac{1}{c^2}
     \log \frac{e^2(M^2)}{e^2(s)} \left( 1- \frac{1}{3}\frac{e^2}{4 \pi^2}
     \sum_{j=1}^{n_f } Q_j^2 N^j_C \log \frac{M^2}{m_j^2}
     \right) \right\}
     \end{eqnarray}
     \begin{equation}
     w_{\gamma_i}(M^2,m_j^2) = 
     \frac{1}{3} \sum_{j=1}^{n_f} \frac{e_j^2}{4 \pi^2} N^j_C \log \frac{M^2}{m_j^2}
     \end{equation}
Note, that $w_{\gamma_i}(M^2,M^2) =0$. Finally, we have
\begin{equation}
w^\theta_{kl} \left(s,\lambda^2 \right)=\frac{e^2}{8 \pi^2}Q_k Q_l\log \frac{s}{\lambda^2} 
\log \frac{2 p_lp_k}{s} \label{eq:wang}
\end{equation}
for the angular dependent corrections from the soft QED regime.

In distinction to the universal terms, the angular dependent corrections in Eq. (\ref{eq:angr})
are not related to the emission probabilities for soft and/or collinear gauge bosons.
As mentioned above, the fundamental objects in the high energy regime above the scale $M$
are the fields in ${\cal L}_{symm}$ as described in Ref. \cite{habil}. In order to clarify the
method, we will compare the predictions in the next section starting at the one loop level.

\section{Comparison with known results} \label{sec:comp}

In this section we will compare our approach with existing calculations. While the presented
results are known\footnote{At one loop, general results for all SL corrections are given
in Ref. \cite{dp} using the physical SM fields.}, 
it serves to illustrate the method and demonstrates the powerful constraints given by
real soft emissions as well as the overall check of the high energy effective theory.
In particular we would like to emphasize that for the universal corrections the splitting
function approach of Refs. \cite{m1,m3} predicts that the factorization
of DL and SL terms for fermions and scalars takes place with the same electroweak group factor
$ - \frac{1}{2} \left( \frac{g^2}{16 \pi^2} T (T+1)
+ \frac{{g^\prime}^2}{16 \pi^2} \frac{Y^2}{4} \right)$.
Only the Yukawa terms and transverse gauge boson anomalous dimensions 
factorize differently, namely with $-\frac{g^2}{32 \pi^2} \frac{3}{2} \frac{m_t^2}{M^2}$ and 
$\frac{g^2}{8 \pi^2} \beta_0 \left(\frac{{g^\prime}^2}{8 \pi^2} \beta^\prime_0 \right)$ correspondingly.

The results of Ref. \cite{bddms} were obtained using the physical fields.
Soft real photon radiation will be included in the comparison.
In the following, the lower index on the cross section indicates the helicity of the electron, where
$e^-_-$ denotes the left handed electron.
We summarize the relevant
results for $e^+_- e^-_+ \longrightarrow W^+_{\rm T} W^-_{\rm T}$,
$e^+_+ e^-_- \longrightarrow W^+_{\rm L} W^-_{\rm L}$ and $e^+_- e^-_+ \longrightarrow
W^+_{\rm L}
W^-_{\rm L}$ from Ref. \cite{bddms} for convenience as follows:
\begin{eqnarray}
&& \!\!\!\!\!\!\!\!\!\!\! \left( \frac{ d \sigma}{d \Omega} \right)_{-,{\rm T}} \!\!\! \approx \! 
\left( \frac{d \sigma}{d \Omega}
\right)^{\rm \!\! Born}_{-,{\rm T}} \!\! \left\{ 1 + \frac{e^2}{8\pi^2} \left[ - 
\frac{1+2c_{\rm w}^2+8c_{\rm w}^4}{
4 c_{\rm w}^2 s_{\rm w}^2} \log^2 \frac{s}{M^2}
+ 3 \frac{1-2c_{\rm w}^2+4c_{\rm w}^4}{4 c_{\rm w}^2 s_{\rm w}^2} \log \frac{s}{M^2}
\right. \right. \nonumber \\ && \;\;\;\;\;\;\;\;\;\;\;
+\frac{4u+2s}{s_{\rm w}^2u} \log \frac{s}{M^2} \log \frac{s}{-t}- \frac{2(1-2c_{\rm w}^2)}{s_{\rm w}^2}
\log \frac{s}{M^2} \log \frac{u}{t} + 3 \log \frac{s}{m_e^2} \nonumber \\ && \;\;\;\;\;\;\;\;\;\;\;
\left. \left.
+ 2 \log \frac{4 \Delta E^2}{s} \!\! \left( \log \frac{s}{m_e^2} + \log \frac{s}{
M^2}+2 \log \frac{t}{u} -2 \! \right) 
- \frac{4}{3} \sum_{j=1}^{n_f} Q_j^2 N^j_C \log \frac{m_j^2}{M^2} \right] \! \right\} 
\label{eq:mt} \\
&& \!\!\!\!\!\!\!\!\!\!\! \left( \frac{ d \sigma}{d \Omega} \right)_{-,\,{\rm L}} \!\!\! \approx \!
\left( \frac{d \sigma}{d \Omega
}
\right)^{\rm Born}_{-,\,{\rm L} } \!\! \left\{ \! 1 \! + \frac{e^2}{8\pi^2} \! \left[ - \frac{1-2c_{\rm w}^2+4
c_{\rm w}^4}{
2 c_{\rm w}^2 s_{\rm w}^2} \log^2 \! \frac{s}{M^2}
+ \frac{103-158 c_{\rm w}^2+80 c_{\rm w}^4}{12 c_{\rm w}^2 s_{\rm w}^2} \log \frac{s}{M^2}
\right. \right. \nonumber \\ && \;\;\;\;\;\;\;\;\;\;\;
- \frac{3 m_t^2}{2 s_{\rm w}^2 M^2} \log \frac{s}{m_t^2}+\frac{4c_{\rm w}^2}{s_{\rm w}^2} \log
\frac{s}{M^2} \log \frac{s}{-t}+ \frac{(1\!-\!2c_{\rm w}^2)^2}{c_{\rm w}^2s_{\rm w}^2}
\log \frac{s}{M^2} \log \frac{u}{t}+3 \log \frac{s}{m_e^2}
\nonumber \\ && \;\;\;\;\;\;\;\;\;\;\; \left. \left.
+ 2 \log \frac{4 \Delta E^2}{s} \!\! \left( \log \frac{s}{m_e^2} + \log \frac{s}{
M^2}+2 \log \frac{t}{u} -2 \! \right) 
- \frac{4}{3} \sum_{j=1}^{n_f} Q_j^2 N^j_C \log \frac{m_j^2}{M^2} \right] \! \right\} 
 \label{eq:ml} \\
 && \!\!\!\!\!\!\!\!\!\!\! \left( \frac{ d \sigma}{d \Omega} \right)_{+,\,{\rm L}}
 \!\!\! \approx \! \left( \frac{d \sigma}{d \Omega
 }
 \right)^{\rm Born}_{+,\,{\rm L}} \! \left\{ \! 1 \!+ \frac{e^2}{8\pi^2} \left[ - \frac{5-10c_{\rm w}^2+8
 c_{\rm w}^4}{
 4 c_{\rm w}^2 s_{\rm w}^2} \log^2 \frac{s}{M^2} + \frac{65-65 c_{\rm w}^2+18 c_{\rm w}^4}{6 c_{\rm
 w}^2 s_{\rm w}^2} \log \frac{s}{M^2}
 \right. \right. \nonumber \\ && \;\;\;\;\;\;\;\;\;\;\;
 - \frac{3 m_t^2}{2 s_{\rm w}^2 M^2} \log \frac{s}{m_t^2}+ \frac{2(1-2c_{\rm w}^2)}{c_{\rm w}^2}
 \log \frac{s}{M^2} \log \frac{u}{t}+3 \log \frac{s}{m_e^2}
  \nonumber \\ && \;\;\;\;\;\;\;\;\;\;\; \left. \left. 
+ 2 \log \frac{4 \Delta E^2}{s} \!\! \left( \log \frac{s}{m_e^2} + \log \frac{s}{
M^2}+2 \log \frac{t}{u} -2 \! \right) 
- \frac{4}{3} \sum_{j=1}^{n_f} Q_j^2 N^j_C \log \frac{m_j^2}{M^2} \right] \! \right\} 
  \label{eq:pl}
  \end{eqnarray}
  where at high energies we denote $t=-\frac{s}{2} \left( 1-\cos \theta \right)$ and
  $u=-\frac{s}{2} \left( 1+\cos \theta \right)$. The angle $\theta$ is the one between the incoming
  $e^+$ and the outgoing $W^+$.
  The Born cross sections are given by:
  \begin{eqnarray}
  \left( \frac{d \sigma}{d \Omega}
  \right)^{\rm Born}_{-,{\rm T}} &=& \frac{e^4}{64 \pi^2s} \frac{1}{4 s_{\rm w}^4} \frac{u^2+t^2}{t^2}
  \sin^2 \theta \label{eq:bmt} \\
  \left( \frac{d \sigma}{d \Omega} \right)^{\rm Born}_{-,\,{\rm L}} &=& \frac{e^4}{64 \pi^2s}
  \frac{1}{16 s_{\rm w}^4 c_{\rm w}^4} \sin^2 \theta \label{eq:bmp} \\
  \left( \frac{d \sigma}{d \Omega} \right)^{\rm Born}_{+,\,{\rm L}} &=& \frac{e^4}{64 \pi^2s}
  \frac{1}{4 c_{\rm w}^4} \sin^2 \theta \label{eq:bpp}
  \end{eqnarray}
  In Eq. (\ref{eq:bmt}) a sum over the two transverse polarizations of the $W^\pm$ ($++$ and $--$)
  is implicit.
  These expressions demonstrate that the longitudinal cross sections
  in Eqs. (\ref{eq:bmp}) and (\ref{eq:bpp}) are not mass suppressed (while $\left( \frac{d \sigma}{d \Omega}
  \right)^{\rm \!\! Born}_{+,{\rm T}}$ is).
  Eqs. (\ref{eq:mt}), (\ref{eq:ml}) and (\ref{eq:pl}) were of course calculated in terms of the physical fields
  of the broken theory and in the on-shell scheme. We denote $c_{\rm w}=\cos \theta_{
  \rm w}$ and $s_{\rm w}=\sin \theta_{\rm w}$ respectively.

In the above results we have included soft bremsstrahlung contributions. In order to demonstrate how
the soft angular dependent real QED corrections serve to fix the virtual angular terms at
the one loop level (and in general also the higher order terms) we list the one loop
purely real corrections separately. To SL accuracy we have for the above cross sections
in the limit $k_0 \leq \Delta E \leq M$:
\begin{eqnarray}
\left( \frac{d \sigma}{d \Omega}
\right)^{\rm \!\! brems}  \!\!\!\!\! &\approx&
\left( \frac{d \sigma}{d \Omega}
\right)^{\rm \!\! Born}  \left(-\frac{e^2}{4 \pi^2} \right) \left\{ 4 \log \frac{2 \Delta E}{\lambda}
-2 \log \frac{2 \Delta E}{\lambda} \log \frac{s}{m_e^2} + 4 \log \frac{2 \Delta E}{\lambda}
\log \frac{u}{t} \right. \nonumber \\ && \left.
-2 \log \frac{2 \Delta E}{\lambda} \log \frac{s}{M^2} + \frac{1}{2} \log^2 \frac{s}{m_e^2} -
\log \frac{s}{m_e^2}+ \frac{1}{2} \log^2 \frac{s}{M^2}-\log \frac{s}{M^2} \right\} \label{eq:re}
\end{eqnarray}
It can be seen in Eq. (\ref{eq:re}) that all angular dependent soft terms are proportional to
$\log \frac{2 \Delta E}{\lambda}$, which means that there is no freedom for the virtual angular
terms in order to avoid the infrared divergence as $\lambda \rightarrow 0$. Note also that all soft
bremsstrahlung corrections are not sensitive to the spin of the final state as expected (we 
use the equivalence theorem such that the longitudinal degrees of freedom are described by
scalar particles).

  Using $e= \frac{g g^\prime}{\sqrt{g^2+{g^\prime}^2}}$, $s_{\rm w}=
  \frac{g^\prime}{\sqrt{g^2+{g^\prime}^2}}$ and $c_{\rm w}= \frac{g}{\sqrt{g^2+
  {g^\prime}^2}}$ we see that the Born cross section
  in Eq. (\ref{eq:bmt}) is proportional to $g^4$, 
  in Eq. (\ref{eq:bmp}) proportional to $(g^2+{g^\prime}^2)^2$ and Eq. (\ref{eq:bpp})
  proportional to ${g^\prime}^4$.
  Thus, in the transverse sector
  the coupling renormalization
  above the scale $M$ is given by:
  \begin{eqnarray}
  g^2(s) &=& g^2(M^2) \left( 1- \frac{g^2(M^2)}{4\pi^2} \left( \frac{11}{12} C_A
  - \frac{1}{24} n_h - \frac{n_{gen}}{3}
  \right) \log \frac{s}{M^2} \right) \nonumber \\ 
  &=& \frac{e^2_{\rm eff} (M^2)}{s_{\rm w}^2} \left( 1 - \frac{e^2_{\rm eff} (M^2)}{
  4 \pi^2 s_{\rm w}^2} \frac{19}{24} \log \frac{s}{M^2} \right)
  \label{eq:rc}
  \end{eqnarray}
  where in the second line we use $C_A$=2, $n_{gen}=3$ and $n_h=1$.
  Below the scale where non-Abelian effects enter, the running is only due to the electromagnetic coupling
  with
  \begin{equation}
  e^2_{\rm eff}(M^2)=e^2 \left(1+ \frac{1}{3} \frac{e^2}{4 \pi^2} \sum_{j=1}^{n_f} Q_j^2 N^j_C 
  \log \frac{M^2}{m_j^2} \right) \label{eq:aeff}
   \end{equation}
   with $\frac{e^2}{4 \pi} =\frac{1}{137}$. The form of Eq. (\ref{eq:aeff}) is purely perturbative
   and does not correctly reproduce the non-perturbative hadronic contribution.
   We observe that the running coupling terms proportional to $\log \frac{s}{M^2}$
   cancel for the process involving transverse gauge bosons with the subleading contributions
   from the virtual splitting functions at one loop
   (see Eq. (\ref{eq:WgRG})) and what remains are just the
   Abelian terms up to scale $M$.

  The RG-corrections to both longitudinal cross sections accounting for the running couplings
  from $M^2$ to $s$
  are given by:
  \begin{eqnarray}
  && \left( \frac{d \sigma}{d \Omega} \right)^{\rm RG}_{-,\,{\rm L}} =
  \left( \frac{d \sigma}{d \Omega} \right)^{\rm Born}_{-,\,{\rm L}} \left\{ 1 +
  \frac{e^2}{8 \pi^2} \frac{41-82 c^2_{\rm w} + 22 c^4_{\rm w}}{6 s^2_{\rm w}c^2_{\rm w}} \log
  \frac{s}{M^2} \right\} \label{eq:mlrg} \\
  && \left( \frac{d \sigma}{d \Omega} \right)^{\rm RG}_{+,\,{\rm L}} =
  \left( \frac{d \sigma}{d \Omega} \right)^{\rm Born}_{+,\,{\rm L}} \left\{1+
  \frac{e^2}{8 \pi^2} \frac{41}{6 c^2_{\rm w}} \log \frac{s}{M^2} \right\} \label{eq:plrg}
  \end{eqnarray}
In order to account for the angular dependent parts in Eqs. (\ref{eq:mt}), (\ref{eq:ml})
and (\ref{eq:pl}) we need to write the amplitude with isospin rotated Born 
matrix elements (in 
the symmetric basis above the scale $M$) as corrections proportional to the physical
Born amplitude. This is always possible since we are dealing with functions of the invariants
only in the high energy limit. In this way
we have, using a regulator mass $M \geq m_i$, the following contributions
relative to the Born amplitude:
\begin{eqnarray}
\sum_{B,W^a} \delta^\theta_{e^+_+e^-_- \longrightarrow W^+_{\rm T} W^-_{\rm T}} &=&
-\frac{g^2}{8\pi^2} \log \frac{s}{M^2} \left( \log \frac{t}{u} + \left(1-\frac{t}{u}\right) \log
\frac{-t}{s} \right) \label{eq:amt} \\ 
\sum_{B,W^a} \delta^\theta_{e^+_+e^-_- \longrightarrow \phi^+ \phi^-} &=&
-\frac{g^2}{8\pi^2} \log \frac{s}{M^2} \left(  \frac{1}{2c_{\rm w}^2}
\log \frac{t}{u} + 2 c_{\rm w}^2 \log
\frac{-t}{s} \right) \label{aml} \\ 
\sum_{B,W^a} \delta^\theta_{e^+_-e^-_+ \longrightarrow \phi^+ \phi^-} &=&
-\frac{{g^\prime}^2}{8\pi^2} \log \frac{s}{M^2} 
\log \frac{t}{u}
\label{apl}   
\end{eqnarray}
In addition we have the soft angular contributions which are given by
\begin{equation}
\sum_{kl} w^\theta_{kl}(s,\lambda) =
-\frac{e^2}{4\pi^2} \log \frac{s}{\lambda^2} 
\log \frac{t}{u}
\label{eq:sv}   
\end{equation}
which is again independent of the spin plus the corresponding matching term (see Eq. (\ref{eq:angr})).
\begin{table}
\begin{center}
\begin{Large}
\begin{tabular}{|l|c|c|r|}
\hline
& T & Y & Q \\
\hline
$e^-_-$ & 1/2 & -1 & -1 \\
\hline
$e^-_+$ & 0 & -2 & -1 \\
\hline
$e^+_+$ & 1/2 & 1 & 1 \\
\hline
$e^+_-$ & 0 & 2 & 1 \\
\hline
$u_-$ & 1/2 & 1/3 & 2/3 \\
\hline
$u_+$ & 0 & 4/3 & 2/3 \\
\hline
$d_-$ & 1/2 & 1/3 & -1/3 \\
\hline
$d_+$ & 0 & -2/3 & -1/3 \\
\hline
$W^\pm_{\rm T}$ & 1 & 0 & $\pm$1 \\
\hline
$\phi^\pm$ & 1/2 & $\pm$1 & $\pm$1 \\
\hline
$\chi$ & 1/2 & $+$1 & 0 \\
\hline
$H$ & 1/2 & $+$1 & 0 \\
\hline
\end{tabular}
\end{Large}
\end{center}
\caption{The quantum numbers of various particles in the electroweak theory.
The indices indicate the helicity of the electrons and quarks. In the high energy
regime described by the Lagrangian in the symmetric basis, we neglect all mass terms, i.e. we
consider all particles as chiral eigenstates with well defined total weak isospin (T)
and weak hypercharge (Y) quantum numbers (except for the photon and the
Z-boson as discussed in the text). In each case, the electric
charge $Q$, measured in units of the proton charge, by the
Gell-Mann-Nishijima formula $Q=T^{3}+Y/2$.
For longitudinally
polarized gauge bosons, the associated scalar Goldstone bosons describe the DL 
and SL asymptotics.}
\label{tab:qn}
\end{table}
  The Sudakov corrections to both cross sections from the infrared evolution equation
  method according to Eq. (\ref{eq:angr}) in the soft photon approximation are given below. The
  quantum numbers are those of the particle-indices and are summarized in Tab. \ref{tab:qn}.
  \begin{eqnarray}
  \left( \frac{ d \sigma}{d \Omega} \right)_{-,{\rm T}}\!\! &=& \!\! \left( \frac{d \sigma}{d \Omega}
  \right)^{\rm Born}_{-,{\rm T}} \left\{ 1 - \left( \frac{g^2}{8 \pi^2} T_{\rm w} (T_{\rm w}+1)
  + \frac{{g^\prime}^2}{8 \pi^2} \frac{Y^2_{\rm w}}{4} \right) \log^2 \frac{s}{M^2} \right. \nonumber \\
  \!\! && \!\! - \left( \frac{g^2}{8 \pi^2} T_{e^-_-} (T_{e^-_-}+1) + \frac{{g^\prime}^2}{8 \pi^2} 
  \frac{Y^2_{e^-_-}}{4}
  \right) \left( \log^2 \frac{s}{M^2}- 3 \log \frac{s}{M^2} \right) \nonumber \\ 
  \!\! && \!\!
-\frac{g^2}{4\pi^2} \log \frac{s}{M^2} \left( \log \frac{t}{u} + \left(1-\frac{t}{u}\right) \log
\frac{-t}{s} \right)  
  - \frac{e^2}{8 \pi^2} \times \nonumber \\
  \!\! && \!\! \left[ \left( \log \frac{s}{m_e^2} - 1 \right) 2 \log \frac{m_e^2}{\lambda^2}
  +\log^2 \frac{s}{m_e^2}-3 \log \frac{s}{m_e^2}- \log^2 \frac{s}{M^2}+3 \log \frac{s}{M^2}\right.
  \nonumber \\ \!\! && \!\! + 2 \left( \log \frac{s}{M^2}-1 \right) \log \frac{M^2}{\lambda^2} - \left(
  \log \frac{s}{m_e^2} - 1 \right) \left( 2 \log \frac{4 (\Delta E)^2}{\lambda^2}-2 \log \frac{s}{m_e^2} \right)
  \nonumber \\ \!\! && -2 \left( \log \frac{s}{M^2}-1 \right)
  \left( \log \frac{4 (\Delta E)^2}{\lambda^2} - \log \frac{s}{M^2} \right) - \log^2 \frac{s}{m_e^2} -
  \log^2 \frac{s}{M^2} \Bigg]
  \nonumber \\
  \!\! && \left. +\frac{e^2}{2 \pi^2} \log \frac{4 (\Delta E)^2}{M^2} \log \frac{t}{u}
  + \frac{2}{3} \frac{e^2}{4 \pi^2} \sum_{j=1}^{n_f} Q_j^2 N^j_C \log 
  \frac{M^2}{m_j^2} \right\}
  \nonumber \\
  \!\! &=& \!\! \left( \frac{d \sigma}{d \Omega}\right)^{\rm \!\! Born}_{-,{\rm T}} \!
  \left\{ 1 - \frac{e^2}{8 \pi^2} \left[ \frac{1+10 c_{\rm w}^2}{4 s_{\rm w}^2 c_{\rm w}^2}
  \log^2 \frac{s}{M^2}
  - 3 \frac{1+2 c_{\rm w}^2}{4 s_{\rm w}^2 c_{\rm w}^2}
  \log \frac{s}{M^2} + \frac{2}{s_{\rm w}^2} \log \frac{s}{M^2} \times \right. \right. \nonumber \\ 
 \!\! && \!\! 
 \left( \left(2 c^2_{\rm w}-1 \right) \log \frac{t}{u} + \left(1-\frac{t}{u} \right) \log
\frac{-t}{s}  \right) 
  -2 \log^2 \frac{s}{M^2} + 3 \log \frac{m_e^2}{M^2} 
  \nonumber \\ \!\!&&\!\! \left. + 4 \log
  \frac{s}{4 (\Delta E)^2} \left( \log \frac{s}{m_e M}+\log \frac{t}{u}-1 \right) \right] 
  \left. + \frac{2}{3} \frac{e^2}{4 \pi^2} \sum_{j=1}^{n_f} Q_j^2 N^j_C \log 
  \frac{M^2}{m_j^2} \right\}
  \label{eq:mymt} 
  \end{eqnarray}
which reproduces the correct result of Eq. (\ref{eq:mt}). For longitudinal degrees of freedom we
use the equivalence theorem and find analogously:
  \begin{eqnarray}
  \left( \frac{ d \sigma}{d \Omega} \right)_{-,{\rm L}}\!\! &=& \!\! \left( \frac{d \sigma}{d \Omega}
  \right)^{\rm Born}_{-,{\rm L}} \left\{ 1 - \left( \frac{g^2}{8 \pi^2} T_{\phi} (T_{\phi}+1)
  + \frac{{g^\prime}^2}{8 \pi^2} \frac{Y^2_{\phi}}{4} \right) \left( \log^2 \frac{s}{M^2}
  -4 \log \frac{s}{M^2} \right) \right. \nonumber
  \\
  \!\! && \!\! - \left( \frac{g^2}{8 \pi^2} T_{e^-_-} (T_{e^-_-}+1) + \frac{{g^\prime}^2}{8 \pi^2}
  \frac{Y^2_{e^-_-}}{4}
  \right) \left( \log^2 \frac{s}{M^2}- 3 \log \frac{s}{M^2} \right)
  \nonumber \\ \!\! && \!\!
-\frac{g^2}{4\pi^2} \log \frac{s}{M^2} \left(  \frac{1}{2c_{\rm w}^2}
\log \frac{t}{u} + 2 c_{\rm w}^2 \log
\frac{-t}{s} \right) \nonumber \\  \!\!&&\!\!
  - 3 \frac{g^2}{16 \pi^2} \frac{m_t^2}{M^2} \log \frac{s}{m_t^2} - \frac{e^2}{8 \pi^2}
  \left[ \left( \log \frac{s}{m_e^2} - 1 \right) 2 \log \frac{m_e^2}{\lambda^2} \right. \nonumber \\
  \!\! && \!\!
  +\log^2 \frac{s}{m_e^2}-3 \log \frac{s}{m_e^2}- \log^2 \frac{s}{M^2}+3 \log \frac{s}{M^2}
  \nonumber \\ \!\! && \!\! + 2 \left( \log \frac{s}{M^2}-1 \right) \log \frac{M^2}{\lambda^2} - \left(
  \log \frac{s}{m_e^2} - 1 \right) \left( 2 \log \frac{4(\Delta E)^2}{\lambda^2}-2 \log \frac{s}{m_e^2}
  \right)
  \nonumber \\ \!\! && -2 \left( \log \frac{s}{M^2}-1 \right)
  \left( \log \frac{4(\Delta E)^2}{\lambda^2} - \log \frac{s}{M^2} \right) - \log^2 \frac{s}{m_e^2} -
  \log^2 \frac{s}{M^2} \Bigg]
  \nonumber \\
  \!\! && \left. +\frac{e^2}{2 \pi^2} \log \frac{4 (\Delta E)^2}{M^2} \log \frac{t}{u}
  + \frac{2}{3} \frac{e^2}{4 \pi^2} \sum_{j=1}^{n_f} Q_j^2 N^j_C \log \frac{M^2}{m_j^2}
  \right\}
  \nonumber \\
  \!\! &=& \!\! \left( \frac{d \sigma}{d \Omega}\right)^{\rm \!\! Born}_{-,{\rm L}} \!
  \left\{ 1 - \frac{e^2}{8 \pi^2} \left[ \frac{1+2 c_{\rm w}^2}{2 s_{\rm w}^2 c_{\rm w}^2}
  \log^2 \frac{s}{M^2}
  - 7 \frac{1+2 c_{\rm w}^2}{4 s_{\rm w}^2 c_{\rm w}^2}
 \log \frac{s}{M^2} + \frac{2}{s_{\rm w}^2}\log \frac{s}{M^2} \times \right. \right. \nonumber \\
 \!\!&&\!\! \left( \frac{(1-2c_{\rm w}^2)^2}{2c_{\rm w}^2} \log \frac{t}{u} + 2 c_{\rm w}^2 \log
 \frac{-t}{s} \right) +\frac{3 m_t^2}{2 s^2_{\rm w} M^2} \log \frac{s}{m_t^2}
  - 2 \log^2 \frac{s}{M^2} + 3 \log \frac{m_e^2}{M^2} \nonumber \\ \!\! && \!\! \left. \left.+ 4 \log
   \frac{s}{4(\Delta E)^2} \left( \log \frac{s}{m_e M}+\log \frac{t}{u}-1 \right) \right]
 + \frac{2}{3} \frac{e^2}{4 \pi^2} \sum_{j=1}^{n_f} Q_j^2 N^j_C \log \frac{M^2}{m_j^2}
  \right\}
 \label{eq:myml}
 \end{eqnarray}
Adding Eqs. (\ref{eq:mlrg}) and (\ref{eq:myml}) yields exactly the result in Eq. (\ref{eq:ml}) from
Ref. \cite{bddms}. Analogously, we have for right handed electrons:
\begin{eqnarray}
\left( \frac{ d \sigma}{d \Omega} \right)_{+,{\rm L}}\!\! &=& \!\! \left( \frac{d \sigma}{d \Omega}
\right)^{\rm Born}_{+,{\rm L}} \left\{ 1 - \left( \frac{g^2}{8 \pi^2} T_{\phi} (T_{\phi}+1)
+ \frac{{g^\prime}^2}{8 \pi^2} \frac{Y^2_{\phi}}{4} \right) \left( \log^2 \frac{s}{M^2}
-4 \log \frac{s}{M^2} \right) \right. \nonumber
\\
\!\! && \!\! - \! \left(\! \frac{g^2}{8 \pi^2} T_{e^-_+} (T_{e^-_+}+1) + \frac{{g^\prime}^2}{8 \pi^2}
\frac{Y^2_{e^-_+}}{4} \!
\right) \!\! \left( \!\log^2 \frac{s}{M^2}- 3 \log \frac{s}{M^2}\! \right) \!
 -3 \frac{g^2}{16 \pi^2} \frac{m_t^2}{M^2} \log \frac{s}{m_t^2} \nonumber \\ \!\! && \!\!
-\frac{{g^\prime}^2}{4\pi^2} \log \frac{s}{M^2} 
\log \frac{t}{u}
- \frac{e^2}{8 \pi^2}
\left[ \left( \log \frac{s}{m_e^2} - 1 \right) 2 \log \frac{m_e^2}{\lambda^2} \right. \nonumber \\
\!\! && \!\!
+\log^2 \frac{s}{m_e^2}-3 \log \frac{s}{m_e^2}- \log^2 \frac{s}{M^2}+3 \log \frac{s}{M^2}
\nonumber \\ \!\! && \!\! + 2 \left( \log \frac{s}{M^2}-1 \right) \log \frac{M^2}{\lambda^2} - \left(
\log \frac{s}{m_e^2} - 1 \right) \left( 2 \log \frac{4 (\Delta E)^2}{\lambda^2}-2 \log \frac{s}{m_e^2}
\right)
\nonumber \\ \!\! && -2 \left( \log \frac{s}{M^2}-1 \right)
\left( \log \frac{4 (\Delta E)^2}{\lambda^2} - \log \frac{s}{M^2} \right) - \log^2 \frac{s}{m_e^2} -
\log^2 \frac{s}{M^2} \Bigg]
\nonumber \\
  \!\! && \left. +\frac{e^2}{2 \pi^2} \log \frac{4 (\Delta E)^2}{M^2} \log \frac{t}{u}
+ \frac{2}{3} \frac{e^2}{4 \pi^2} \sum_{j=1}^{n_f} Q_j^2 N^j_C \log \frac{M^2}{m_j^2}
 \right\}
\nonumber \\
 \!\! &=& \!\! \left( \frac{d \sigma}{d \Omega}\right)^{\rm \!\! Born}_{+,{\rm L}} \!
\left\{ 1 - \frac{e^2}{8 \pi^2} \left[ \frac{5-2 c_{\rm w}^2}{4 s_{\rm w}^2 c_{\rm w}^2}
 \log^2 \frac{s}{M^2}
 -  \frac{4- c_{\rm w}^2}{s_{\rm w}^2 c_{\rm w}^2}
 \log \frac{s}{M^2} + \frac{2}{c_{\rm w}^2} \log \frac{s}{M^2} \times \right. \right. \nonumber \\
\!\! && \!\! \left( 1-2c_{\rm w}^2 \right) \log \frac{t}{u}-
2 \log^2 \frac{s}{M^2} + 3 \log \frac{m_e^2}{M^2} 
+\frac{3 m_t^2}{2 s^2_{\rm w} M^2} \log \frac{s}{m_t^2} \nonumber \\
  \!\! && \!\! \left. \left.+ 4 \log
   \frac{s}{4(\Delta E)^2} \left( \log \frac{s}{m_e M}+\log \frac{t}{u}-1 \right) \right]
+ \frac{2}{3} \frac{e^2}{4 \pi^2} \sum_{j=1}^{n_f} Q_j^2 N^j_C \log \frac{M^2}{m_j^2}
 \right\}
 \label{eq:mypl}
\end{eqnarray}
Again we see that after adding Eqs. (\ref{eq:plrg}) and (\ref{eq:mypl}) we obtain the result in
Eq. (\ref{eq:pl}) from
Ref. \cite{bddms}. Thus we have demonstrated that to subleading logarithmic accuracy our results
from the infrared evolution equation method in conjunction with the Goldstone boson equivalence
theorem are
identical with existing one loop calculations with physical fields in the high energy limit.

In a similar fashion, at the one loop level all DL and SL results from the infrared evolution
equation method agrees with the general one loop corrections obtained from the physical fields
in Ref. \cite{dp}. 

At higher orders, only for the process of massless $e^+e^-\longrightarrow {\overline f} f$
production results are available in Ref. \cite{kps} to SL and in Ref. \cite{kmps} even to SSL order.

The Born amplitude in the high energy limit for $f \neq e$ is given by
\begin{equation}
{\cal M}^{\rm Born}_{e^+_{\alpha} e^-_{\alpha} \longrightarrow {\overline f}_{\beta} f_{\beta}}
= \frac{i}{s} \left[ g^2 T^3_{e^-_{\alpha}} T^3_{f_\beta} + {g^\prime}^2 \frac{Y_{e^-_{\alpha}} Y_{f_\beta}}{4}
\right]  \langle e^+, \alpha| \gamma_\mu | e^-, \alpha \rangle 
\langle {\overline f}, \beta| \gamma^\mu | f, \beta \rangle
\end{equation}

In the high energy regime we have at one loop order for the angular corrections
relative to the Born amplitude:
\begin{eqnarray}
\sum_{B,W^a} \delta^\theta_{e^+_{\alpha} e^-_{\alpha} \longrightarrow {\overline f}_{\beta} f_{\beta}} 
\!\!&=&\!\!\!
-\frac{g^2}{16\pi^2} \log \frac{s}{M^2} \left\{ \left[ \tan^2 \theta_{\rm w} Y_{e^-_{\alpha}} Y_{f_\beta}
+ 4 T^3_{e^-_{\alpha}} T^3_{f_\beta} \right] \log \frac{t}{u} \right. \nonumber \\ && \!\!\!
\left. + \frac{\delta_{\alpha, -} \delta_{\beta,-}}{\tan^2 \theta_{\rm w} Y_{e^-_{\alpha}} Y_{f_\beta} /4
+ T^3_{e^-_{\alpha}} T^3_{f_\beta}} \left( \delta_{d,f} \log \frac{-t}{s} - \delta_{u,f}
\log \frac{-u}{s} \right) \!\! \right\} \label{eq:efang}
\end{eqnarray}
where the last line only contributes for left handed ($-$) fermions and the $d,u$ symbols denote the
corresponding isospin quantum number of $f$.
In addition we have the soft angular contributions which are given by
\begin{equation}
\sum_{k,l} w^\theta_{kl}(s,\lambda) =
-\frac{e^2}{4\pi^2} Q_e Q_f \log \frac{s}{\lambda^2} 
\log \frac{t}{u}
\label{eq:svf}   
\end{equation}
and the corresponding matching term with $\lambda \rightarrow M$.
Subtracting Eq. (\ref{eq:svf}, $\lambda \rightarrow M$) from Eq. 
(\ref{eq:efang}) is in agreement with Eq. (50) of Ref. \cite{kps}.

Also the higher order SL terms from Eq. (\ref{eq:angr}) are in agreement with the results of
Ref. \cite{kps} (up to yukawa enhanced terms which are neglected in a massless
theory). The importance of these and their phenomenology is discussed in
Refs. \cite{kps,kmps,habil}.

\section{Conclusions} \label{sec:con}

In this paper we have put forth the arguments allowing for a consistent SL resummation of logarithmically
enhanced terms in the SM. Our strategy uses as an input only the knowledge of analogous terms
in unbroken massless non-Abelian theories and in QED (including mass terms). 
While DL and universal SL terms have already
been treated in earlier publications all the arguments given given here for the angular terms
can also be understood as a general justification for the matching approach, since the
constraints stemming from the real soft QED regime apply for all terms.
Here we have focussed on the non-universal process dependent
angular corrections originating from the exchange of the electroweak gauge bosons.

We have argued that spontaneous symmetry breaking has effectively no bearing on these
particular terms. We have used the infrared finiteness of semi-inclusive cross sections to argue
that the mass gap of the electroweak gauge bosons does not lead to new effects at higher orders
since the soft real QED corrections fix all possible virtual terms containing $\log \lambda$.
Similarly, all fermion mass singularities are fixed by the KLN theorem and in particular, are due
only to photon exchange assuming $m_i \leq M$. This is also true for novel CKM mixing effects
in the massive quark sector.

From direct inspection of the Feynman rules we have concluded that no Yukawa dependent angular terms
arise which are not mass suppressed. On the other hand, the scalar sector yields angular terms at
higher orders in analogy to a scalar non-Abelian theory after the application of the equivalence
theorem. 

The soft (CKM-extended) isospin correlations are included in matrix form.
This is meant to apply to those external lines for which an isospin
can be assigned. For the photon and the $Z$-boson, the notation applies to the symmetric basis
amplitudes only. The result can be formulated in exponentiated operator form in 
the (CKM-extended) isospin space.

We have shown how to apply our results at the one loop level and find agreement with all known
results. Also at higher orders we agree with the known corrections for massless four fermion
processes.

In summary, we have shown that a full effective SL description is possible based on the concept
of the infrared evolution equation method. The accompanying soft 
QED corrections can all be accommodated
for through the appropriate matching conditions at the weak scale.

\section*{Acknowledgments}

I would like to thank A.~Denner for helpful discussions and M.~Spira for
carefully reading the manuscript.

\end{document}